\documentclass[10pt]{article}
\usepackage{amsmath}
\usepackage{amssymb}
\usepackage{graphicx}
\usepackage[dvips]{color}
\usepackage{dcolumn}
\usepackage{bm}

\begin{document}

\begin{figure}
\leftline{\includegraphics[scale=0.5]{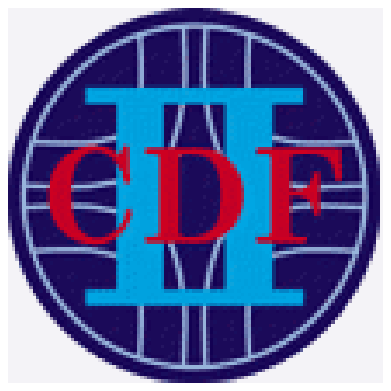}\hfill
FERMILAB-CONF-04-149-E }
\end{figure}


\title{  Pentaquark Searches at CDF }
\author{ 
 Igor V. Gorelov \thanks{talk given on behalf of the CDF
 Collaboration at the XII International Workshop on Deep Inelastic
 Scattering,~DIS~2004,~14~-~18~April~2004,
 $\rm{\check{S}trbsk\acute{e}}$ Pleso, High Tatras, Slovakia. To be published in Proceedings of ~DIS~2004.} \\
 (For the CDF Collaboration) \\ 
 \small{\textit{Department of Physics and Astronomy,}}\\ 
 \small{\textit{University of New Mexico,}}\\ 
 \small{\textit{800 Yale Blvd. NE, Albuquerque, NM 87131, USA}}\\
 \small{\textit{email:gorelov@fnal.gov}} 
 }
\date{}
\maketitle

\begin{abstract}
  Experimental results of a search for the $\Xi_{3/2}(1860)$ cascade
  pentaquark state in data collected with the CDF 2 Detector in Run II
  at the Tevatron are presented. No evidence for these states in the
  neutral $\Xi^{-}\pi^{+}$ and doubly charged $\Xi^{-}\pi^{-}$ modes
  has been found. Preliminary upper limits on yields at 1862~MeV/c$^2$
  relative to the well established resonance $\Xi^{*}(1530)^{0}$ are
  presented.
\end{abstract}



\section{Introduction}	
 Evidence for an exotic pentaquark state $\Theta^{+}(1540)$ with
 strangeness $S=+1$ has been claimed by a number of experimental
 groups. Enhancements with a significance of 4.4 to 7.0 standard
 deviations have been observed in the invariant mass of $K^{+}N$ in
 photoproduction \cite{thetapl:kn} and in $pK^{0}_{s}$
 \cite{thetapl:pk0}. The signals seen have been assigned to a
 $\Theta^{+}(1540)$ state. These results have inspired a search for
 other exotic baryon states. The NA49 Collaboration has
 reported~\cite{xi:na49} the observation of a strangeness $S=-$2,
 isospin $I=$3$/$2 state $\Xi^{--}_{3/2}\rightarrow\Xi^{-}\pi^{-}$.
 An indication of a neutral mode in $M(\Xi^{-}\pi^{+})$ has been
 demonstrated~\cite{xi:na49} as well. Recently the H1 Collaboration at
 HERA has published \cite{thetac:h1} an observation of a narrow
 anti-charmed baryon state in the mode $D^{*+}\bar{p}$ at
 $\sim$3099~MeV/c$^{2}$ and interpreted this as a heavy pentaquark
 $\Theta^{0}_{c}$.
\par 
 The pentaquark state $\Theta^{+}$, according to the chiral soliton
 model~\cite{theory:soliton}, is considered as a bound state of five
 quarks. Experimental evidence for pentaquark $\Theta^{+}$ suggested
 the existence of other pentaquark partners classified within the
 antidecuplet $\overline{10}$ representation~\cite{theory:soliton} or
 the $\overline{10}_{f}\oplus 8_{f}$ multiplet as predicted by the
 constituent quark model approach in \cite{theory:quark}.
\par 
 The experimental status of pentaquark baryons includes some
 controversy.  The signal of $\Theta^{+}$ claimed by
 \cite{thetapl:kn,thetapl:pk0} is not confirmed by
 \cite{theta:notseen}. The cascade pentaquark claimed by
 \cite{xi:na49} has not been seen by \cite{xi:notseen}.  Negative
 results on both $\Theta^{+}$ and $\Xi^{--}_{3/2}$ have been reported
 by large statistics experiments \cite{notseen}.  These experiments
 exploit their excellent mass resolution and large data samples to
 calibrate the mass spectra of interest by well established states
 like $\Lambda(1520)\rightarrow pK^{-}$ and
 $\Lambda^{+}_{c}\rightarrow pK^{0}_{s}$ (for $\Theta^{+}$ searches)
 and $\Xi^{*}(1530)^{0}\rightarrow\Xi^{-}\pi^{+}$ (for $\Xi_{3/2}$
 searches).
\par 
 Recently the CDF Collaboration undertook a comprehensive search for
 several pentaquark states using its Run II 220~pb$^{-1}$ of data
 taken with the upgraded CDF 2 Detector. We present here the
 particular search for the cascade pentaquark $\Xi_{3/2}$ through its
 modes $\Xi^{0}_{3/2}\rightarrow\Xi^{-}\pi^{+}$ and
 $\Xi^{--}_{3/2}\rightarrow\Xi^{-}\pi^{-}$, both of which involve the
 doubly strange cascade baryon $\Xi^{-}$ in the final state. The
 analysis is based on two data samples. The first one was collected by
 a trigger selecting events with at least two tracks of opposite
 charge each, having a momentum above 2.0~GeV/c and an impact
 parameter measured by the CDF silicon detector to be larger than
 100~$\mu$m. The total momentum of both tracks was required to be
 larger than 5.5~GeV/c.  This ``two displaced track trigger'' sample
 is enriched by events with heavy quarks decaying via hadronic
 modes. A complementary dataset was taken with a trigger requiring an
 inclusive jet of transverse energy $E_{T}>$20~GeV.
\section{Cascades in the CDF 2 Detector}
 The final state cascade baryon $\Xi^{-}$ \footnote{Unless otherwise
 stated all references to the specific charge combination imply the
 charge conjugate combination as well.} decays almost 100$\%$ of time
 into $\Lambda^{0}\pi^{-}$, with a subsequent decay of the
 $\Lambda^{0}$. Since the days of bubble chamber experiments this
 spectacular mode has been identified and reconstructed as a single
 track vertex (we call it here \verb!VERTEX_1!) formed by a kinked
 track, presumably the $\pi^{-}$, followed in the fiducial volume by
 another, $V^{0}-$vertex (\verb!VERTEX_2!), presumably the
 $\Lambda^{0}$ decaying to $p\pi^{-}$, as sketched at Fig.1. In our
 analysis \verb!VERTEX_2! was subjected to a 3-dimensional fit.  Then
 the three tracks $p\pi^{-}$ and $\pi^{+,-}$ were fitted to a common
 3-dimensional vertex with the constraint that
 $M(p\pi^{-})=M_{PDG}^{\Lambda^{0}}$ and that \verb!VERTEX_2! points
 back to \verb!VERTEX_1!, see also Fig.1.
\par
\vspace*{0.20cm}
\begin{tabular}{c c}
\begin{minipage}{3.5in}
\raggedright
\includegraphics[height=0.20\textheight]{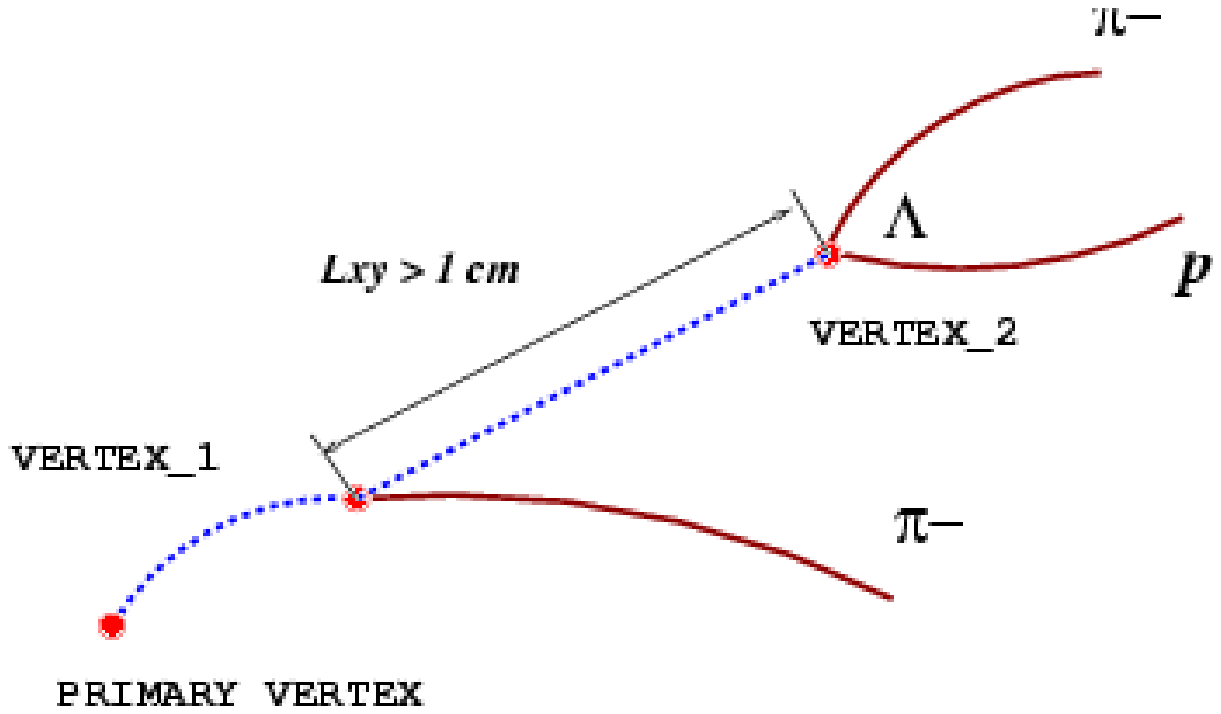}
\end{minipage}
&
\begin{minipage}[h]{2.5in}
\raggedright
\begin{small}
  Figure~1: Sketch of the cascade decay topology.  The $\Lambda^{0}$
  candidates with $M(p\pi^{-})\in M_{PDG}^{\Lambda}\pm$5MeV/c$^{2}$
  were fitted to the vertex and the 2-dimensional
  $\chi^{2}_{r\phi}<49.0$ was required. For $\Xi$ candidates
  $M(p\pi^{-}\pi_{kink})\in$ $M_{PDG}^{\Xi}\pm$60MeV/c$^{2}$ and no
  $p\pi^{-}\pi_{kink}$ vertex fit quality $\chi^{2}$ cut were
  required. The $\Xi$ and $\Lambda$ vertices are separated by more
  than 1~cm in the transverse plane and have an impact parameter
  $d_{0}(\Xi)<$150~$\mu$m defined in the transverse plane as well.
\end{small}
\end{minipage}
\end{tabular}
\par
\vspace*{0.20cm}
 The resulting invariant mass spectrum of cascade candidates
 $M(\Xi^{-}\rightarrow \Lambda^{0}\pi^{-})$ is shown on Fig.2.  The
 clear signal at the $\Xi^{-}$ mass is seen on top of a large
 combinatorial background.
\par
\begin{tabular}{c c}
\begin{minipage}{3.0in}
\raggedright
\includegraphics[height=0.24\textheight]{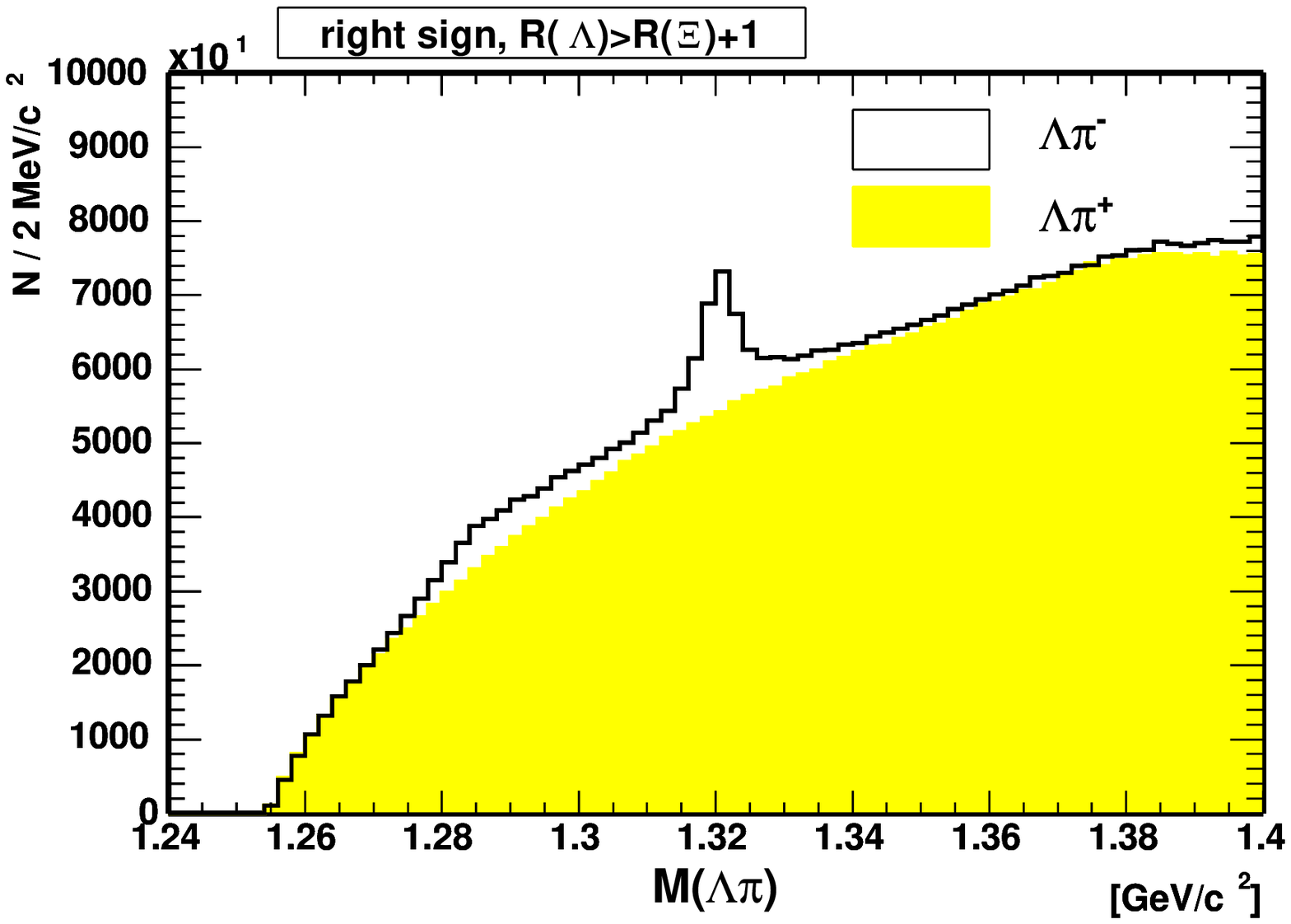}
\end{minipage}
&
\begin{minipage}[h]{3.0in}
\raggedright
\begin{small}
  Figure~2: The invariant mass spectrum of $M(\Lambda^{0}\pi^{-})$
  after vertex fits described above. The cascade signal is present
  with a large combinatorial background. The shaded histogram
  corresponds to the wrong charge combinations
  $M(\Lambda^{0}\pi^{+})$.
\end{small}
\end{minipage}
\end{tabular}
\par
 The long lifetime of $\Xi^{-}$ hyperons ($c\tau$=4.91~cm) permits
 reconstruction of their tracks from hits in the CDF silicon tracker
 (SVX II).  A novel technique developed by CDF uses the vertex
 position and momentum of a cascade hyperon fitted in the CDF outer
 tracker to seed the hyperon tracking in SVX II. This procedure
 results in a substantial background reduction and improved vertex and
 impact parameter resolution of the $\Xi^{-}$, see Fig.~3. The overall
 relative efficiency of the hyperon reconstruction with SVX II hits is
 $\sim$40$\%$. The yields of cascades in the CDF 2 Detector are larger
 by a factor of $\sim$20 than the ones found by the NA49
 experiment~\cite{xi:na49}.
\par
\begin{tabular}{c c}
\begin{minipage}{3.0in}
\raggedright
\includegraphics[height=0.24\textheight]{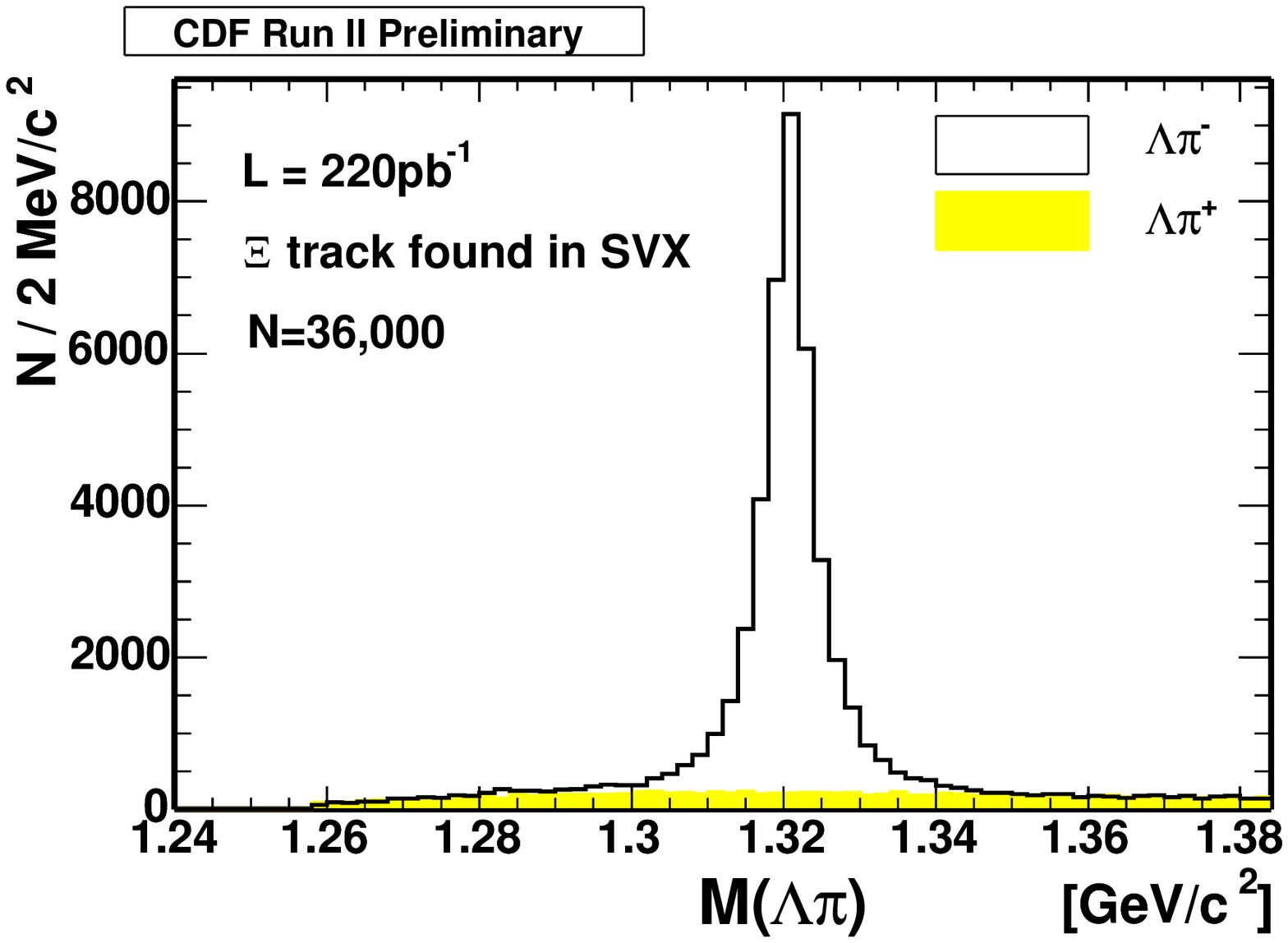}
\end{minipage}
&
\begin{minipage}[h]{3.0in}
\raggedright
\begin{small}
  Figure~3: The invariant mass spectrum of $\Xi^{-}$ hyperons which
  have tracks successfully reconstructed in the silicon tracker. A cut
  on impact parameter $d_{0}(\Xi)<$150~$\mu$m was applied, selecting
  hyperons produced promptly in the primary vertex region. The very
  clean signal based on the data sample of integrated
  $\mathcal{L}$=220~pb$^{-1}$ taken by the two displaced track trigger
  contains $\sim$36000 events. The analogous signal (not shown here)
  for the inclusive jet $E_T>$20.0~GeV dataset of the same
  $\mathcal{L}$ contains $\sim$4870 events.
\end{small}
\end{minipage}
\end{tabular}
\section{Pentaquarks in the $\Xi^{-}\pi^{+}$ and $\Xi^{-}\pi^{-}$ Decay Modes}
 The hyperon tracks reconstructed in SVX II with mass $M(\Xi)\in
 M^{\Xi}_{PDG}\pm10$~MeV/c$^2$ (see Fig.~3) were combined with all
 remaining tracks with $P_{T}>$400~MeV/c and 3 or more hits in the SVX
 II tracker.  Then the track pairs $\Xi^{-}\pi^{+,-}$ were subjected
 to a vertex fit constrained by the requirement for the secondary
 vertex to point to the primary one. The invariant mass spectra
 $M(\Xi^{-}\pi^{+,-})$ are shown in Fig.~4 and 5.
\par
\vspace*{0.15cm}
\begin{tabular}{c c}
\begin{minipage}{3.0in}
\raggedright
\includegraphics[height=0.24\textheight]{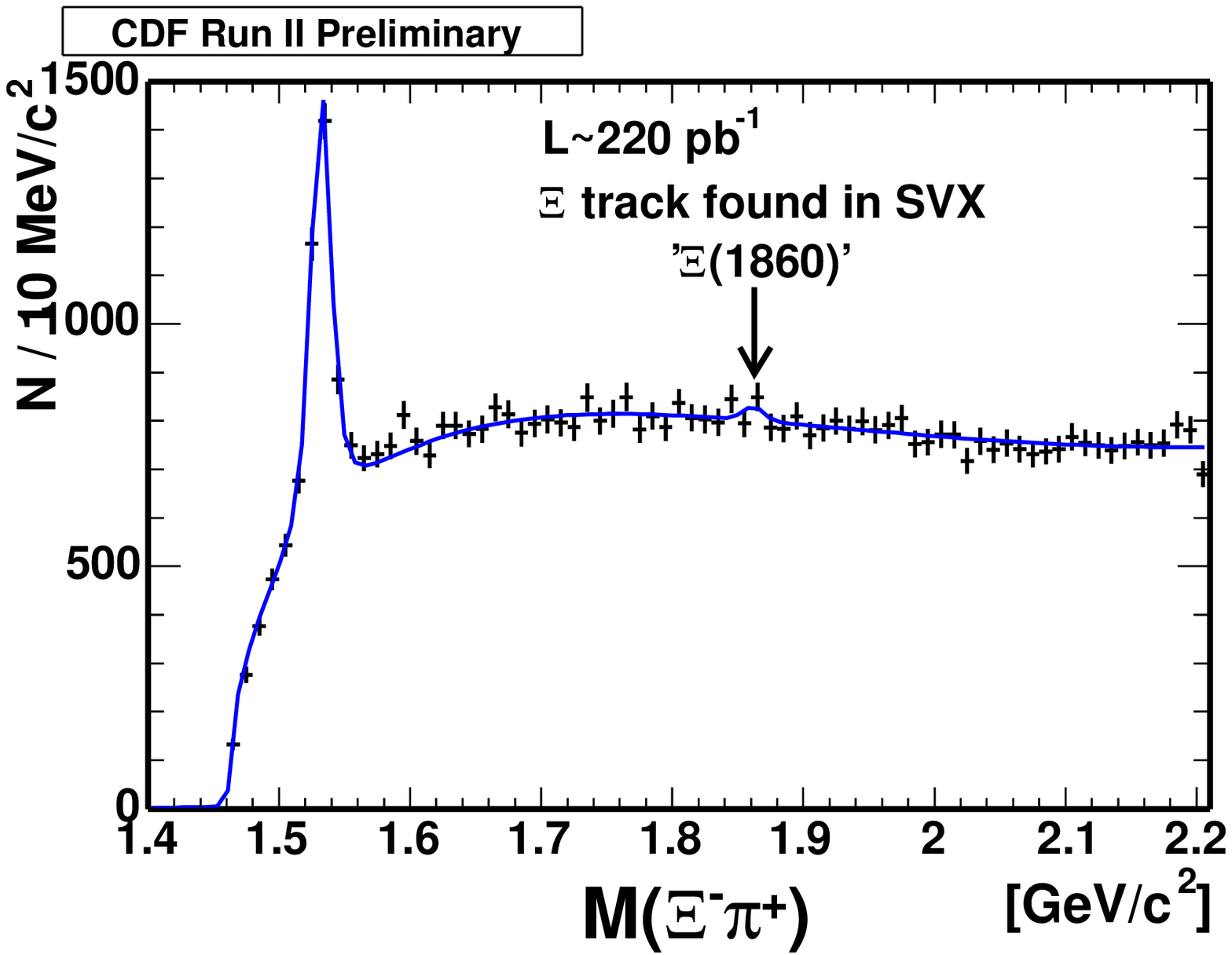}
\end{minipage}
&
\begin{minipage}[h]{3.0in}
\raggedright
\begin{small}
  Figure~4: The invariant mass spectrum for $\Xi^{-}\pi^{+}$ whose
  hyperons have tracks successfully reconstructed in the SVX II. The
  track pairs form the vertex fitted with two-dimensional
  $\chi^{2}_{r\phi}<36.0$. The two displaced track trigger data sample
  of integrated luminosity $\mathcal{L}$=220~pb$^{-1}$ is used. The
  fit to a Breit-Wigner convoluted with a Gaussian finds 2182$\pm$92
  events in the peak for
  $\Xi^{*}(1530)^{0}\rightarrow\Xi^{-}\pi^{+}$. The peak is used as a
  gauge signal for pentaquark searches. The similar spectrum (not
  shown here) in the inclusive jet $E_T>$20~GeV sample yields
  387$\pm$34 events for $\Xi^{*}(1530)^{0}$. The pentaquark signal
  region at 1862~MeV/c$^2$ is fitted with a Gaussian of a fixed width
  $\sigma=$8~MeV/c$^2$ predicted by Monte-Carlo simulation.
\end{small}
\end{minipage}
\end{tabular}
\par
\begin{tabular}{c c}
\begin{minipage}[h]{3.0in}
\raggedright
\includegraphics[height=0.24\textheight]{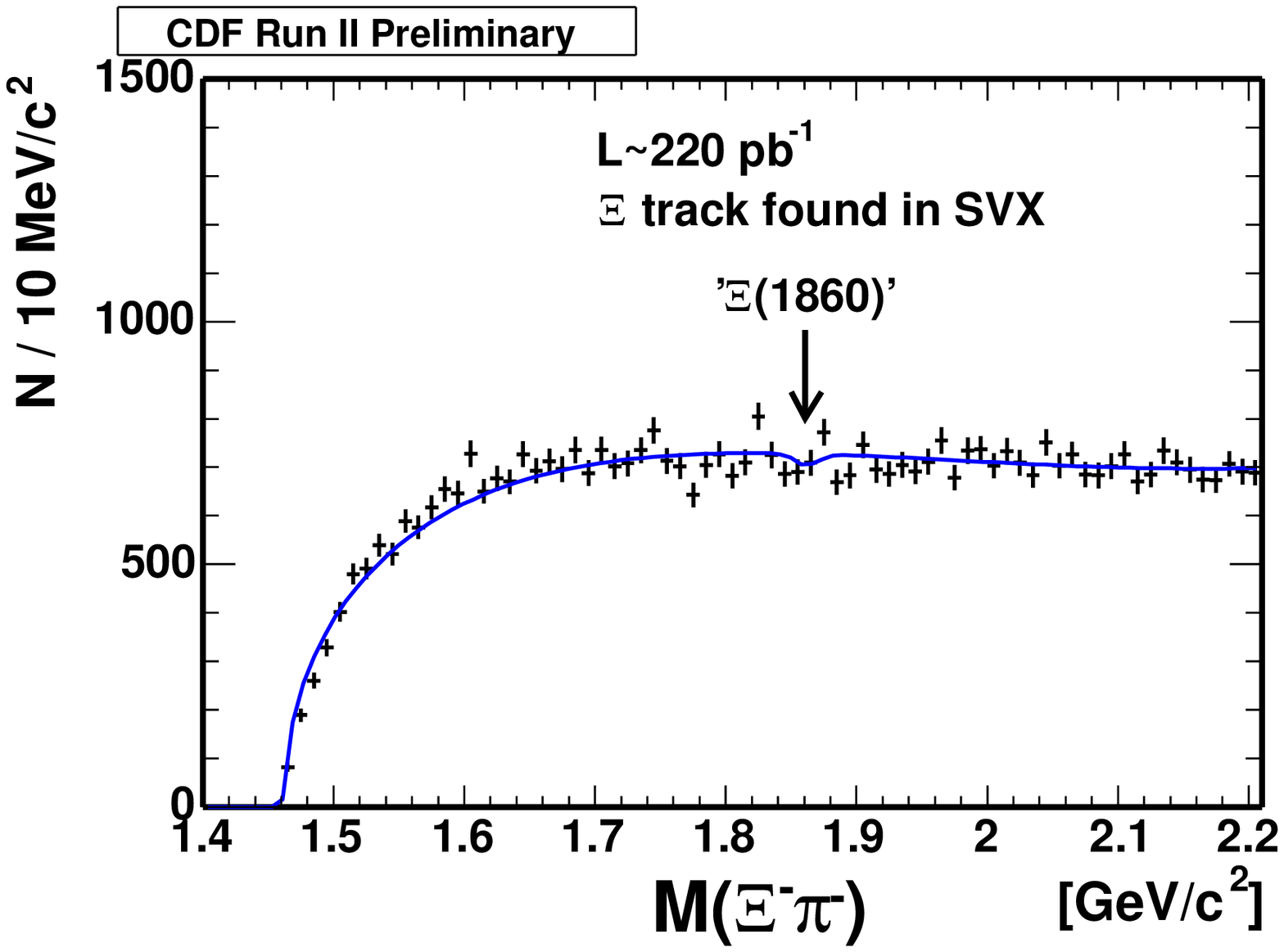}
\end{minipage}
&
\begin{minipage}[h]{3.0in}
\raggedright
\begin{small}
  Figure~5: The invariant mass spectrum for $\Xi^{-}\pi^{-}$ whose
  hyperons have tracks successfully reconstructed in the SVX II. The
  two displaced track trigger data sample of
  $\mathcal{L}$=220~pb$^{-1}$ is used. The spectrum is fitted by a
  polynomial background shaped by a square-root threshold
  function. The NA49 signal region at $M(\Xi^{-}\pi^{-})=$
  1862~MeV/c$^2$ is fitted with a Gaussian of fixed width
  $\sigma=$8~MeV/c$^2$. A similar spectrum (not shown here) is
  observed for the inclusive jet $E_T>$20~GeV sample.
\end{small}
\end{minipage}
\end{tabular}
\par
 The spectra $M(\Xi^{-}\pi^{+})$ (Fig.~4) and $M(\Xi^{-}\pi^{-})$
 (Fig.~5) corresponding to neutral and doubly charged cascade
 pentaquark modes do not reveal any enhancement around
 M=1862~MeV/c$^2$~\cite{xi:na49}.  Similar results have been obtained
 with the inclusive jet $E_T>$20~GeV sample. We have set upper limits
 on the production of pentaquarks decaying via both modes. These are
 shown in a Table 1 below.
\par
\vspace*{0.20cm}
\begin{tabular}{c c}
\begin{minipage}{3.5in}
\raggedright
%
\begin{center}
\begin{tabular}{|c|c|c|}
\hline
Mode       & $@90\% C.L.$ & $@95\% C.L.$ \\
\hline
\multicolumn{3}{|c|}{Two Displaced Track Trigger Sample}  \\
\hline
$\sigma\cdot$Br($\Xi^{-}\pi^{+}$)$/$$\sigma\cdot$Br($\Xi^{*}(1530)^{0}$) & 0.06 & 0.07  \\
$\sigma\cdot$Br($\Xi^{-}\pi^{-}$)$/$$\sigma\cdot$Br($\Xi^{*}(1530)^{0}$) & 0.03 & 0.04  \\
combined statistics                & 0.07 & 0.08  \\
\hline
\multicolumn{3}{|c|}{Inclusive Jet $E_T>$20~GeV Sample}  \\
\hline
$\sigma\cdot$Br($\Xi^{-}\pi^{+}$)$/$$\sigma\cdot$Br($\Xi^{*}(1530)^{0}$) & 0.06 & 0.08  \\
$\sigma\cdot$Br($\Xi^{-}\pi^{-}$)$/$$\sigma\cdot$Br($\Xi^{*}(1530)^{0}$) & 0.07 & 0.09  \\
combined statistics                & 0.09 & 0.11  \\
\hline
\end{tabular}
\end{center}
%
\end{minipage}
&
\begin{minipage}[h]{2.5in}
\raggedright
\begin{small}
  Table~1: Upper limits set for a $\Xi^{0}_{3/2}$ and $\Xi^{--}_{3/2}$
  pentaquark states. The yields were calculated relative to the
  calibrating signal of $\Xi^{*}(1530)^{0}$ seen in both data samples.
\end{small}
\end{minipage}
\end{tabular}
\section{Summary}
 The CDF Collaboration conducted a search for doubly strange $S=-$2
 pentaquark states in the $\Xi^{-}\pi^{+}$ and $\Xi^{-}\pi^{-}$ decay
 modes. The signals of the basic hyperon $\Xi^{-}$ state comprised
 $\sim$36000~events in the two displaced trigger dataset and
 $\sim$4900~events in the inclusive jet $E_T>$20~GeV dataset. The well
 established resonance $\Xi^{*}(1530)^{0}\rightarrow\Xi^{-}\pi^{+}$
 was used as a calibrating signal and yielded 2182$\pm$92 events from
 the two displaced trigger sample and 387$\pm$34 events from the
 inclusive jet $E_T>$20~GeV sample. No evidence of exotic baryon
 states produced in CDF Detector at Tevatron has been found. Upper
 limits on production of states in the mass range of
 $\sim$1862~MeV/c$^2$ have been set. CDF Collaboration is pursuing a
 vigorous program of searches for possible pentaquark production at
 the Tevatron.
\section{Acknowledgments}
 The author is grateful to his colleagues from the CDF Pentaquark
 Working Group, especially to Dr.~D.~Litvintsev for useful suggestions
 and comments made during preparation of this talk. The author would
 like to thank Prof.~Sally~C.~Seidel for support of this work,
 fruitful discussions, and comments.
%

\end{document}